\documentclass[11pt]{article}
\usepackage{moriond,epsfig}

\def\be{\begin{equation}}
\def\ee{\end{equation}}
\def\bea{\begin{eqnarray}}
\def\eea{\end{eqnarray}}

\begin{document}
\vspace*{4cm}
\title{HIGH ENERGY NEUTRINOS FROM THE COLD: \\STATUS AND PROSPECTS OF THE ICECUBE EXPERIMENT}

\author{ C. Portello-Roucelle, for the IceCube collaboration}

\address{Lawrence Berkeley National Laboratory,
One Cyclotron Road,                                                
Berkeley, CA  94720\\}

\maketitle\abstracts{The primary motivation for building neutrino telescopes is to open the road for neutrino astronomy, and to offer another observational window for
the study of cosmic ray origins. Other physics topics, such as the search for WIMPs, can also be developed with neutrino telescope. 
As of March 2008, the IceCube detector, with half of its strings deployed, is the world largest neutrino telescope taking data to date and it will reach its completion in 2011. Data taken with the growing detector are being analyzed. The results of some of these works are summarized here. AMANDA has been successfully integrated into IceCube data acquisition system and continues to accumulate data. Results obtained using only AMANDA data taken between the years 2000 and 2006 are also presented. The future of IceCube and the extensions in both low and high energy regions will finally be discussed in the last section. }

\section{Motivations for neutrino astronomy with IceCube} \label{astro}

We expect the acceleration of cosmic rays in astrophysical objects to be accompanied by the production of high energy neutrinos via $pp$ or $p\gamma$ interactions at the acceleration site~\cite{nuastroPrinciple}. The detection of these neutrinos could provide us with fundamental information about these sources of cosmic rays, the most violent objects in the universe.
The preferred candidates for astro-accelerators are expected to have large-scale strong shocks and/or strong magnetic field, such as active galactic nuclei (AGN), supernovae remnants, microquasars or gamma ray bursts. The study of such objects is on-going in gamma ray astronomy which has produced an impressive harvest of results in the last few years. Moreover, the recent announcement of the Auger collaboration of a possible correlation between the arrival direction of cosmic rays with energies in excess of  6$\times10^{19}$ eV and AGN~\cite{AugerAGN} reinforces the interest in the studies on bottom-up processes for cosmic rays production and multi-messenger studies. For instance, if an AGN origin of the highest energy cosmic rays were to be confirmed, the neutrino flux from those objects could be within reach of a kilometer scale detector such as IceCube within a few years~\cite{NuAuger}. \\

Neutrinos are a completely unique tool for the study of the cosmic ray sources. Charged protons below $10^{18.5}$ eV are bent by the inter-galactic magnetic fields and no longer point back to their sources thus making proton astronomy impossible at low energies. As for neutral messengers, the gamma rays are strongly attenuated above 50 TeV because of their interaction with the infrared background, reaching us only from galactic sources at high energies. The neutrinos, with their weak interaction cross section, are the only particles that allow us to explore the non thermic universe at cosmological distances and at all energies. Nevertheless, if this very small cross section is an advantage for the propagation from the source, the detection of neutrinos from astronomical events requires a very large detection volume. A kilometer cube scale is needed to allow the detection of a measurable number of events in one year from the expected diffuse neutrino flux at the Waxman-Bahcall bound~\cite{WB}.
Several large scale neutrino telescopes are currently taking data or under development using either water (Baikal, ANTARES, NEMO, NESTOR and the future kilometer scale KM3Net) or ice as a detection medium (IceCube and its sub-detector AMANDA). The sensitivity of IceCube to astrophysical neutrino sources is given in a previous article~\cite{sensitivity}. We will give here a brief overview of the results of AMANDA for neutrino astrophysics in section~\ref{AMANDA}, and will present the first point source search with a partial configuration of IceCube in section~\ref{IC9}. In addition to these analyses, IceCube will also look for high energy neutrinos that are expected to be created by interaction of the cosmic ray protons with the background radiation~\cite{GZKnu}. 
More specific studies done with IceCube for supernovae detection~\cite{SN} and GRB searches~\cite{GRB} can also be found in the literature.

\section{Other physics potential of the experiment} \label{plus}
In addition to high energy neutrino astronomy, IceCube's scientific reach extends to particle physics by looking for neutrino from annihilation of weakly interacting massive particles (WIMP), like the neutralino. These dark matter candidates are expected to accumulate in gravitational potential wells such as the Earth or Sun. If the annihilation products of two trapped neutralinos include neutrinos, a neutrino flux excess may be observed in the direction of the center of the Earth or Sun. As a consequence, neutrino telescopes like IceCube, IceCube DeepCore (see section \ref{DeepCore}) and AMANDA aim at the indirect detection of dark matter with the Sun or the Earth as effective neutrino sources~\cite{wimps}. The results obtained with AMANDA are presented in section \ref{AMANDA}.  \\
Other topics of particle physics can also be adressed but will not be discussed here, like the search for magnetic monopoles~\cite{monopole}, strange quark matter or SUSY Q-balls~\cite{Qballs}.    

\section{The IceCube detector}

IceCube is located at the geographic South Pole. Neutrinos are detected through their charged current interactions in the ice of the detector volume (or in the ice surrounding the detector for muon and tau neutrinos). The Cherenkov light produced by the charged lepton resulting from this interaction travels through the transparent ice and is collected by the digital optical modules~\cite{DOM} (DOMs) of IceCube.

\begin{figure}
\center{
\includegraphics[height=2.5in]{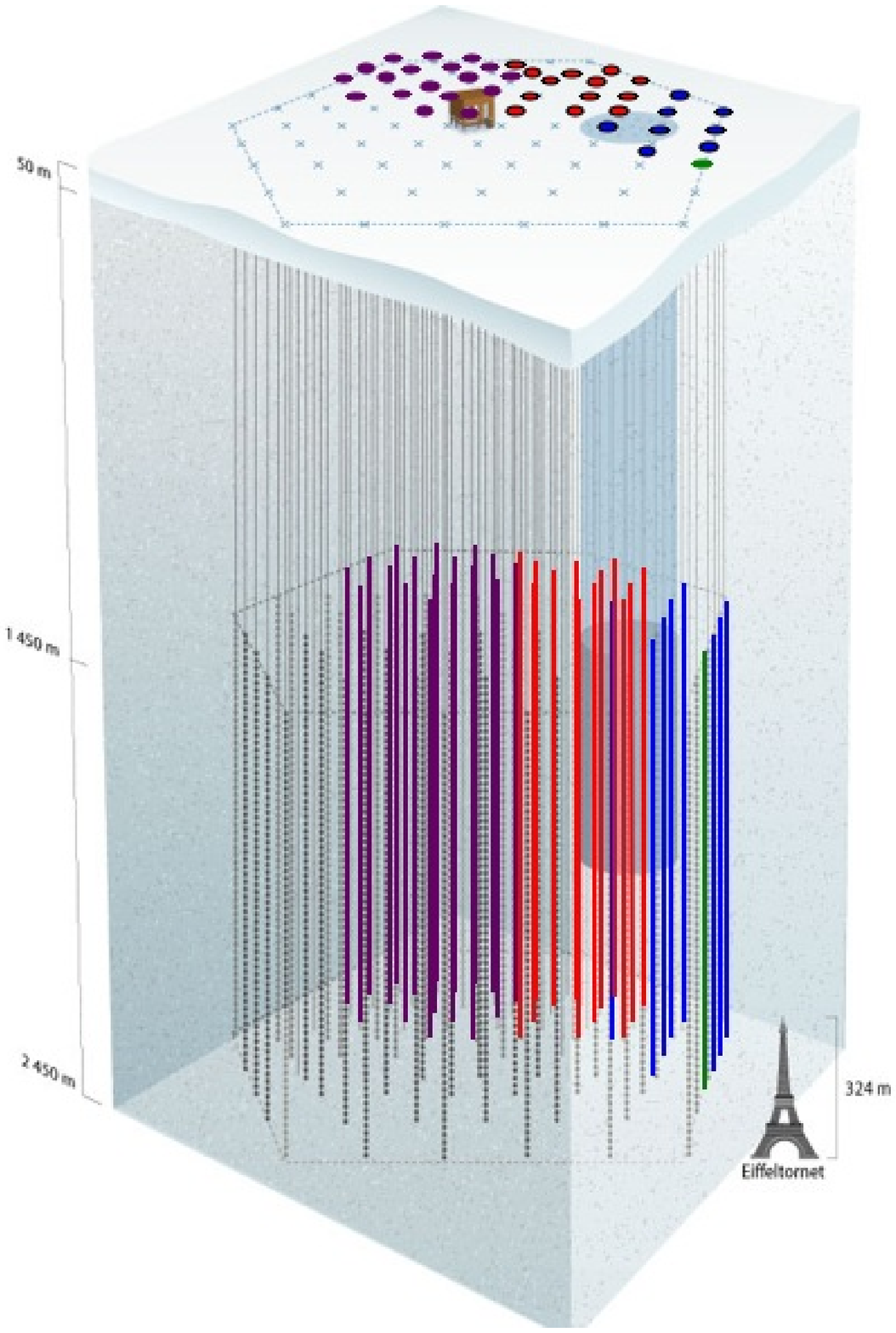}
\hspace{2cm} 
\includegraphics[height=2.5in]{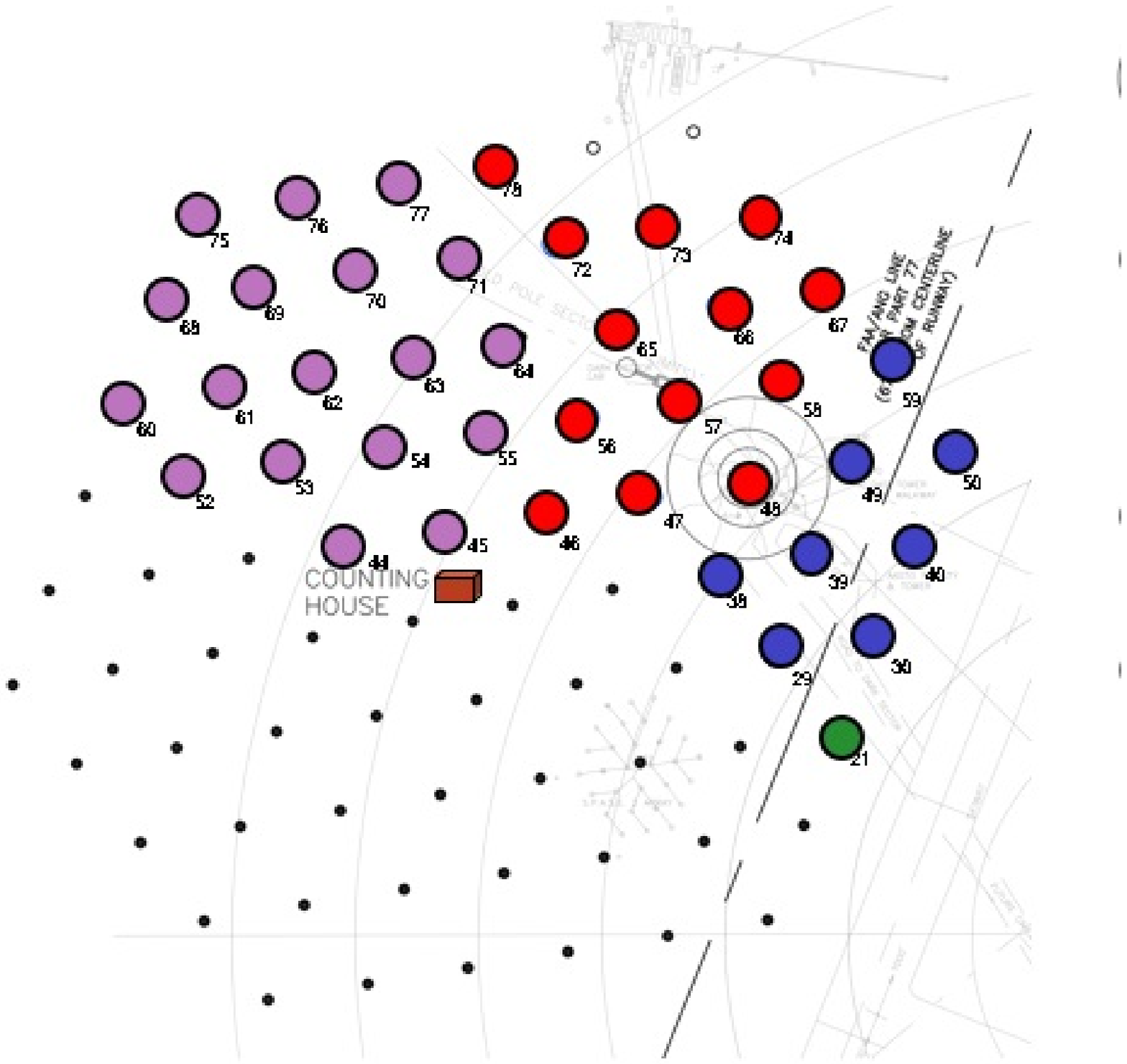}
}
\caption{The IceCube detector side view ({\it on the left}) and top view ({\it on the right}). Currently installed lines (40) in ice are indicated. Current in-ice strings include: one string (in green) deployed in 2005, eight (in blue) in 2006, 13 (in red) in 2006 and 22 (in purple) in 2007. The AMANDA detector appears in the right part of in-ice. The IceTop surface array is also shown.
\label{fig:IC40}}
\end{figure}

The current IceCube design~\cite{design} consists of 80 strings, each bearing 60 DOMs. They are deployed at depths between 1450 and 2450 m below the surface of the ice, forming the in-ice part of the detector. The DOMs have a spacing of 17m on each string and the strings form a triangular grid pattern with an inter-string spacing of 125 m, providing a 1 km$^3$ instrumented volume. The buried detector is topped on the surface by an array of 80 stations called IceTop~\cite{IceTop} for the study of extensive air showers  (see fig.\ref{fig:IC40}). 
Each IceTop station, located above an IceCube string, consists of two tanks filled with ice. Each of those tanks contains two DOMs of same design as the one used for the in-ice part of the detector. The surface array can be operated looking for anti-coincidence with the in-ice events to reject downgoing muons or in coincidence, to provide a usefull tool for cosmic ray composition studies~\cite{HighPt,compo}. 

Each DOM used by IceCube comprises a 10" Hamamatsu R7081-02 photomultiplier tube (PMT) housed in a glass pressure vessel and in situ data acquisition electronics. This electronics is the heart of the IceCube data acquisition system: it reads out, digitizes, processes and buffers the signals from the PMT. When the individual trigger conditions are met at the DOM, it reports fully digitized waveforms to a software-based trigger and event builder on the surface. The electronics acquires in parallel on Analog Transcient Waveform Digitizers (ATWDs) at 300 megasamples per second (MSPS) sampling over a 425 ns window. In addition the electronics also records the signal with a coarser 40 MSPS sampling over a 6.4 $\mu$s window to catch the late part of the signals. Two parallel sets of ATWDs on each DOM operate in alternation so that one is active and ready to acquire while the other is read out. This design greatly reduces the dead-time of an individual DOM. The time calibration yields a timing resolution with a RMS narrower than 2ns for the signal sent by the DOM to the surface~\cite{timeres}. The noise rate due to random hits observed for in-ice DOMs is of the order of 300 Hz. This very low value gives us the possibility to monitor the DOM hit rates and to use it to have a sensitivity to low energy (MeV) neutrinos from supernova core collapse throughout the Milky Way and out to the Large Magellanic Cloud~\cite{SN}. 

IceCube has also integrated its predecessor, the AMANDA detector, as it is now surrounded by IceCube (see fig.~\ref{fig:IC40}). AMANDA consists in 677 analog optical modules distributed on 19 strings with a much denser configuration than IceCube (string spacing of approximately 40m), giving it a lower energy threshold. The AMANDA optical modules are less sophisticated than the IceCube DOMs. The pulse processing electronics and data acquisition system is on the surface and the signal from AMANDA OMs has to be transmitted over roughly 1 km before being treated. Roughly half of the 677 AMANDA OMs transmit their signals to the surface over optical fibers, which allows for a timing accuracy of 2 to 3 ns, comparable to the one of the DOMs, although with greatly reduced dynamic range. The other half of the OMs are connected to the surface only by electrical cables, which stretch the pulses substantially thus separation of successive pulses is prevented. 
For relatively low energy events, the dense configuration of AMANDA gives it a considerable advantage over IceCube. Moreover, IceCube strings surrounding AMANDA can be used as an active veto against cosmic ray muons, making the combined IceCube + AMANDA detector considerably more effective for low energy studies than AMANDA alone. The DeepCore upgrade, whose construction will start next austral summer, will provide IceCube with a dense subdetector using DOM technology. This upgrade will open many possibilities in the low energy region and WIMPs studies as discussed in section~\ref{DeepCore}.  \\

The data taking with the partially finished IceCube detector is running smoothly and the detector is operating as expected.  The detector began taking data in 2006 with a nine strings configuration (IC-9) and with a 22 strings configuration in 2007 (IC-22). The data acquisition with AMANDA also continues, enabling analyses done with more than 7 years of accumulated data. The analysis of the IC-9 configuration of IceCube has already lead to first results with atmospheric neutrinos which are detailed in section \ref{IC9}. The analysis of IC-22 data is on-going and will be finished during the summer 2008.    \\ 

\section{Summary of current AMANDA results} \label{AMANDA}

\subsection{Search for astrophysical sources}
\begin{figure}
\center{\includegraphics[height=1.5in]{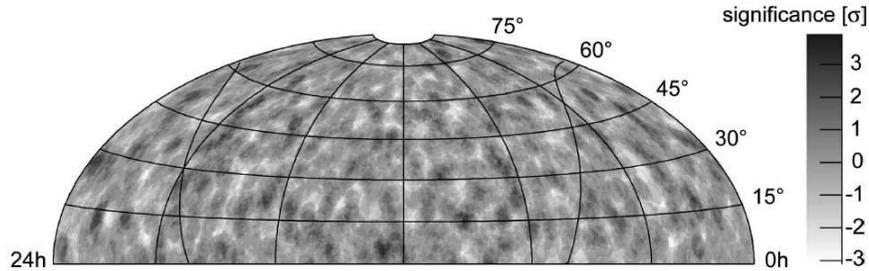}}
\caption{Significance map for the Northern hemisphere sky obtained for the 2000-2004 AMANDA II data analysis. The result obtained is compatible with random fluctuations.   
\label{fig:SigMap}}
\end{figure}

Between 2000 and 2004, AMANDA-II, the final configuration of the AMANDA detector as an independent entity, has been taking data. Results on the 5 years of the dataset have been reported. This subset yields 4282 up-going neutrino candidates with an estimated background contamination of approximately 5\%. The analysis for point sources in the Northern hemisphere sky~\cite{AMANDAps} for this dataset yielded no statistically significant point source of neutrinos as can be seen in Fig.~\ref{fig:SigMap}. The highest positive deviation corresponds to about 3.7$\sigma$. The probability of such a deviation or higher due to background, estimated with 100 equivalent sky surveys of events with
randomized right ascension, is 69\%. Based on these studies, an upper limit has been placed on a reference E$^{-2}$ point source flux of muon neutrinos averaged over declination in the Northern hemisphere sky at 90 \% confidence level: $E^2 d\phi/dE  < 5.5 \times 10^{-8} \rm GeVcm^{-2} s^{-1}$ in the energy range of 1.6 TeV to 2.5 TeV. \\
Over the same period of time, a search for neutrino emission from 32 specific candidate sources chosen based on observations at various wavelengths in the electromagnetic spectrum has been performed~\cite{AMANDAps}. 
No statistically significant evidence for neutrino emission was found from any of the candidate sources. The highest observed significance, with 8 observed events compared to 4.7 expected background events ($1.2\sigma $), is at the location of the GeV blazar 3C273. The second highest excess ($1.1 \sigma$) is from the direction of the Crab Nebula, with 10 observed events compared to 6.7 expected background events. 

In addition to searches for individual sources of neutrinos, AMANDA data taken between 2000 and 2003 have been used to set a limit on possible diffuse fluxes of neutrinos. Populations of distant sources could lead to such a diffuse flux that would clearly prove the acceleration of hadrons in astrophysical sources even if the sources cannot be resolved. This diffuse flux can be distinguished from the background of atmospheric neutrinos due to its harder spectra, expected from most astrophysical sources. This study relies on the the number of triggered OMs which serve as an energy estimator for AMANDA. A limit of $E^{2} d\phi/dE < 7.4 \times10^{-8}~ \rm GeVcm^{-2} s^{-1} sr^{-1}$ is placed on the diffuse muon neutrino flux in the energy range from 16 TeV to 2.5 PeV at 90\% confidence level~\cite{AMANDAdiff}. Additionally, AMANDA has searched for an all-flavour diffuse flux from the Southern sky, a work on these three years of data places a limit of $E^{2} d\phi/dE < 2.7\times 10^{-7} \rm ~GeV cm^{-2}s^{-1}sr^{-1}$, in the energy range of $2\times 10^{5}$ to $10^{9}$ GeV~\cite{Lisa}. 

\subsection{Searches for neutralino dark matter}

\begin{figure}
\begin{center}
\includegraphics[height=2.5in]{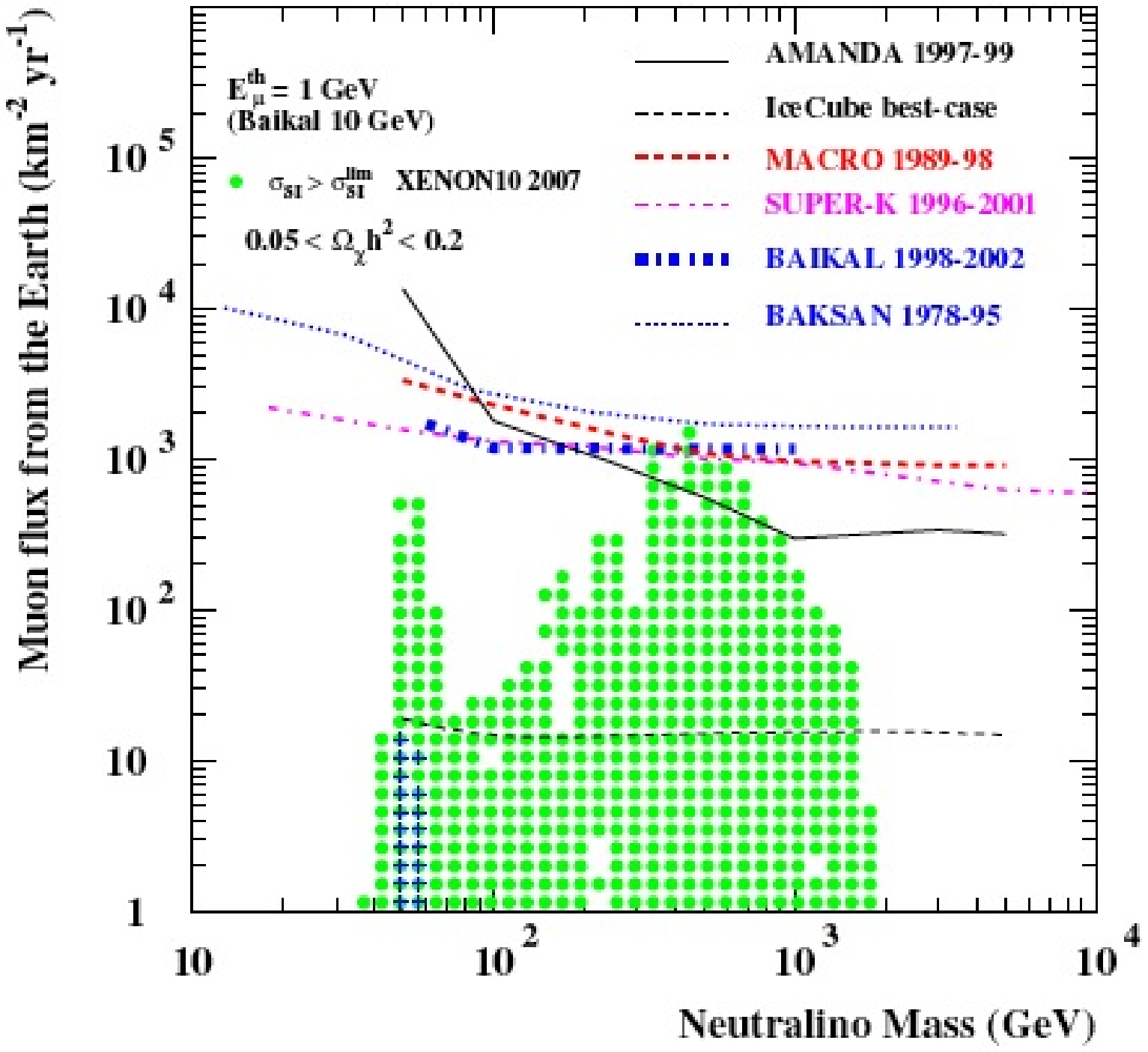}
\hspace{1cm}
\includegraphics[height=2.5in]{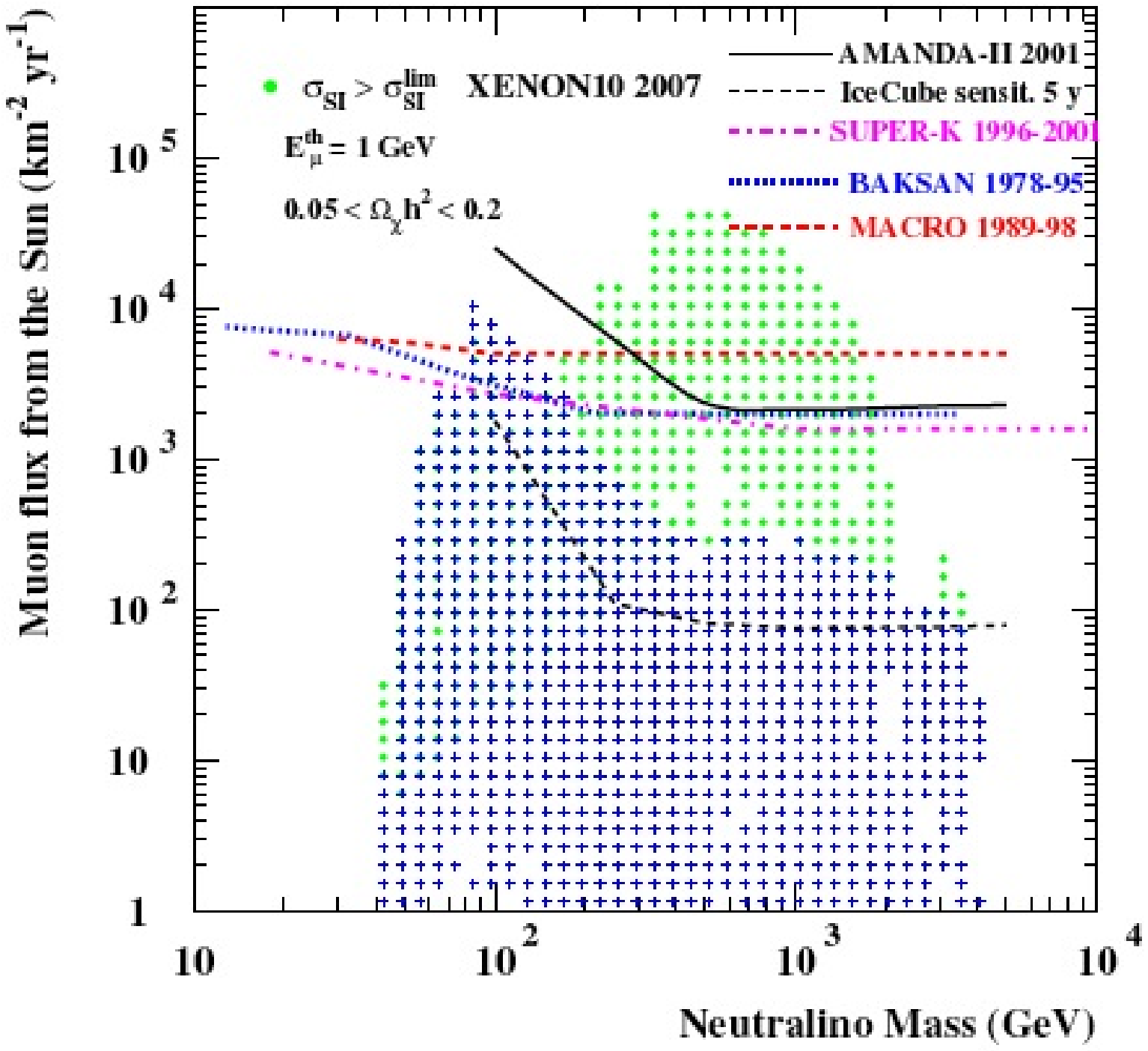}
\end{center}
\caption{90\% CL upper limit on the muon flux from neutralino annihilations in the center of the Earth (left) and from the Sun (right). Markers show predictions for cosmologically relevant MSSM models, the dots representing models excluded by XENON10 \label{fig:wimps}}
\end{figure}

AMANDA can be used to search for neutralino dark matter by looking for a neutrino flux excess from the center of the Earth~\cite{wimpsE} or from the Sun~\cite{wimpsS}. The respective limits obtained with the 2001-2003 dataset for the Earth and the 2001 dataset for the Sun are given in Fig.~\ref{fig:wimps}. The figures show the muon flux limit from neutralino annihilations, along with the results from other indirect searches and predictions from theoretical models. Disfavoured models by recent direct searches with the XENON 10 experiment~\cite{xenon} are shown as green dots.

\section{First results from the IceCube 9 strings configuration}\label{IC9}

The IC-9 dataset has a total livetime of 137.4 days taken between June and November 2006. 234 neutrino candidates were identified on this data sample with 211$\pm$76 (syst.) $\pm$ 14 (stat.) events expected from atmospheric neutrinos and less than 10\% pollution by the background of down-going muons~\cite{IC9Atm}. \\

\begin{figure}
\includegraphics[height=2in]{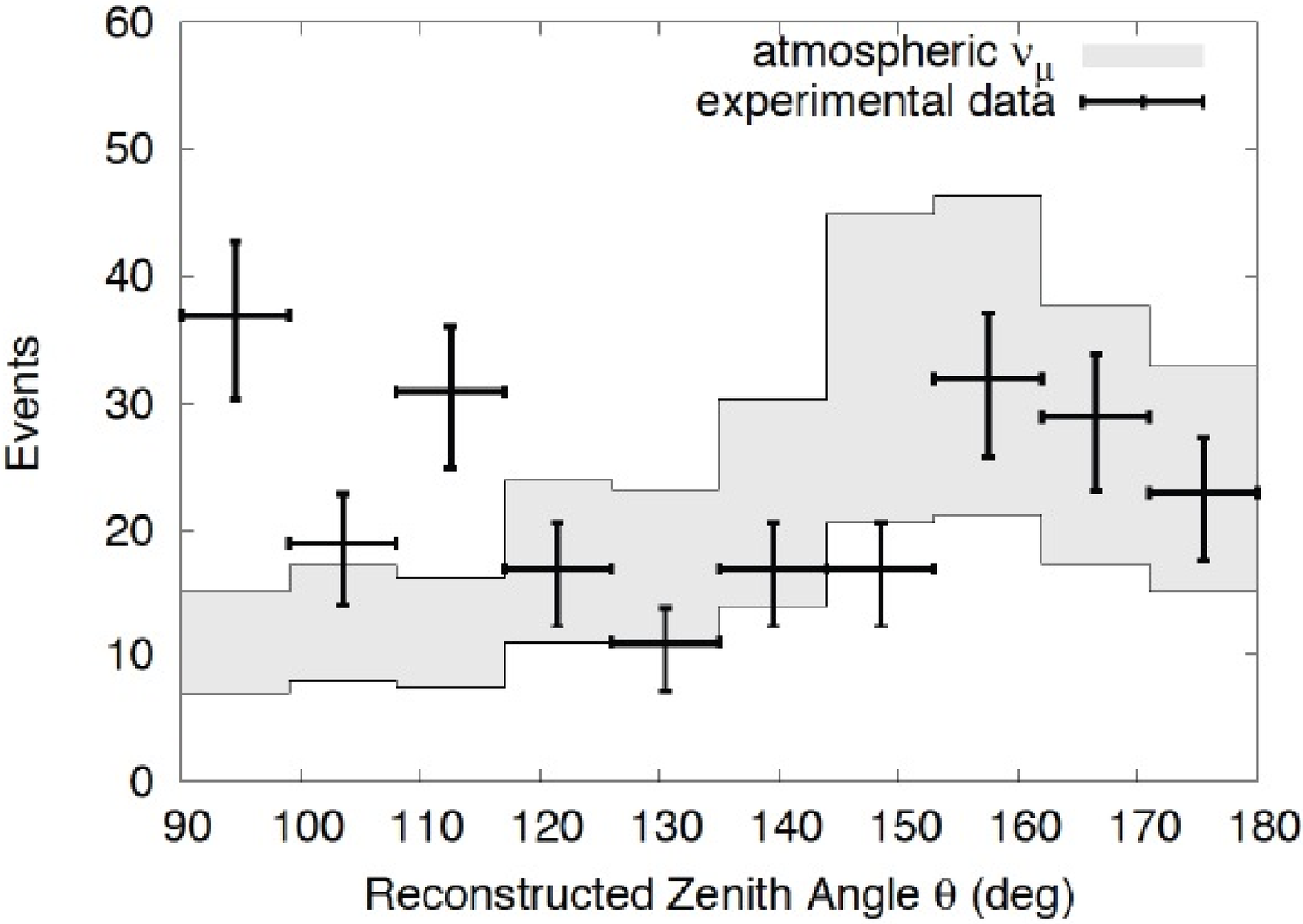}
\includegraphics[height=2in]{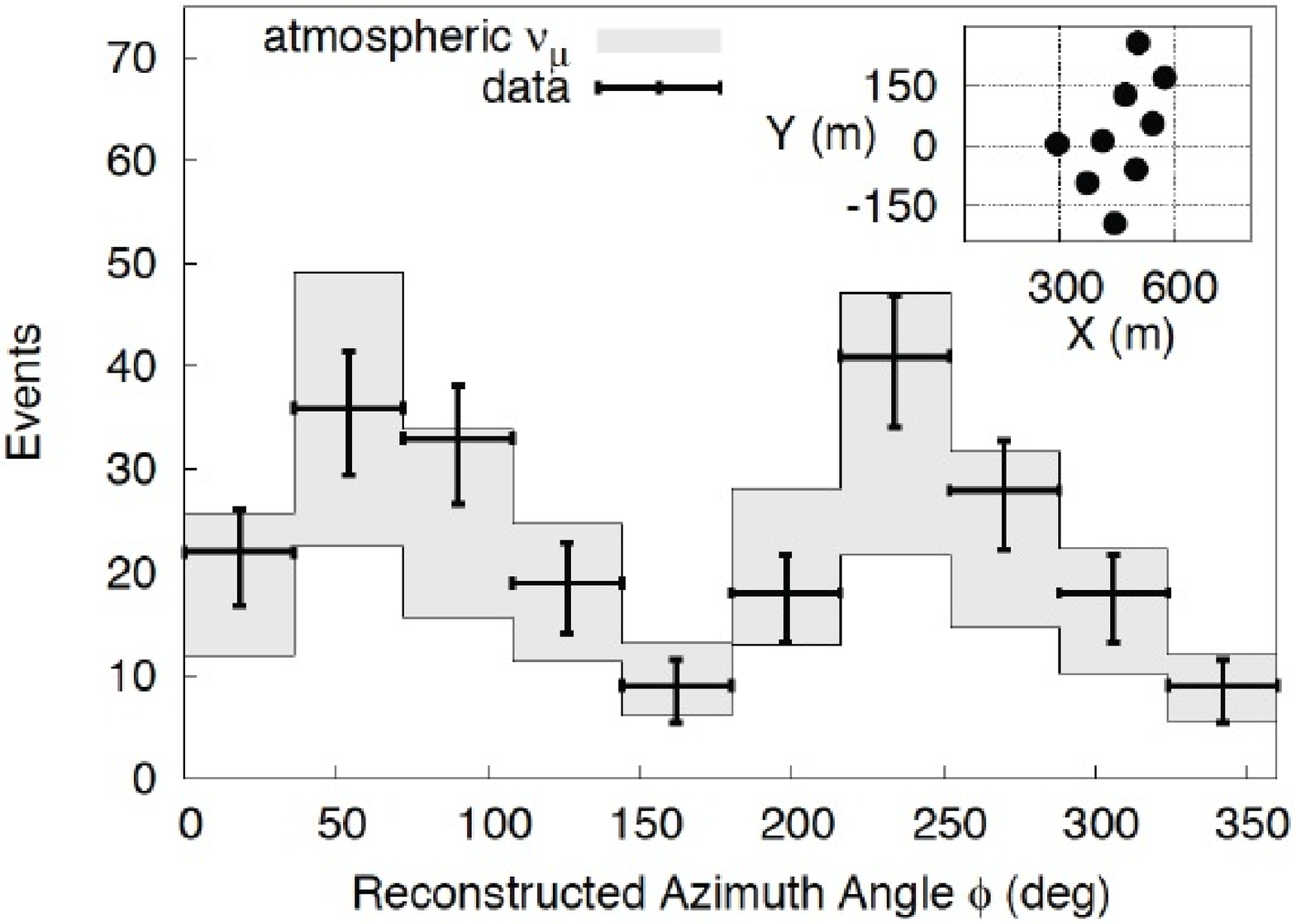}
\caption{Zenith angle distribution (on the left) and azimuth angle distribution (on the right) for the final sample of IC-9 events. A zenith angle of 90 degrees corresponds to an horizontal event; a straight up-going event has a zenith angle of 180 degrees. The shadowed area indicates the simulation expectations with systematic errors. The error bars are statistical only. The configuration of the IceCube string seen from the top is also plotted on the right. The prefered axis of this configuration explains the features observed in the azimuth angle distribution. 
\label{fig:AngDistAtm}}
\end{figure}

The zenith and azimuth angle distributions of these neutrino candidates are shown on Fig.~\ref{fig:AngDistAtm}. The agreement with simulation is good except for a discrepancy near the horizon due to a residual contamination of down going muons. This discrepancy would disappear with tighter event selection. One can notice the 2 strong peaks in the azimuth angle distribution (entry on the right in Fig.~\ref{fig:AngDistAtm} on the right), due to the very asymetric configuration of the detector and corresponding to the long axis of IC-9  . \\ 

\begin{figure}
\begin{center}
\includegraphics[height=2.5in]{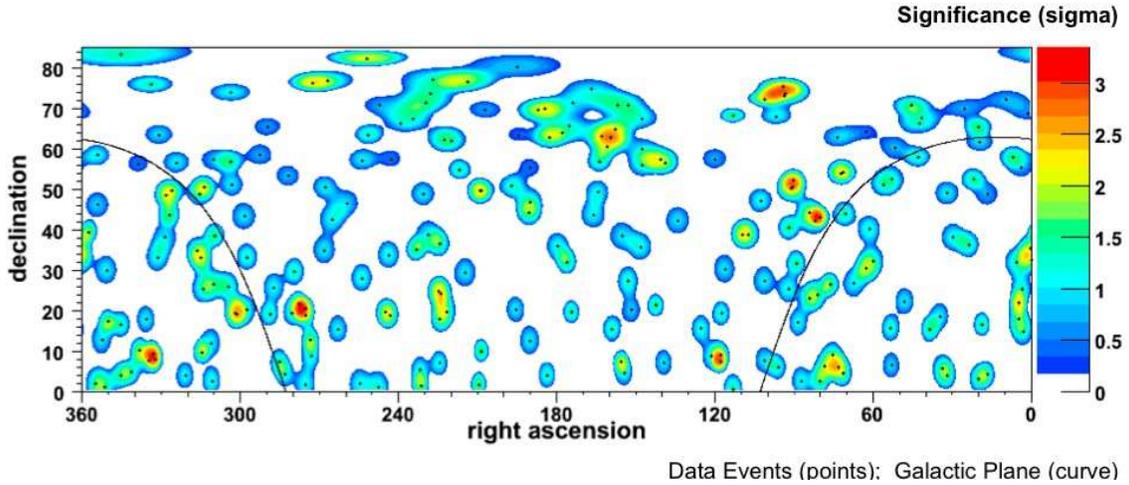}
\end{center}
\caption{Significance map of the Northern hemisphere sky obtained with IC-9.  The excesses observed are consistent with random fluctuations of an uniform background.  
\label{fig:IC9sky}}
\end{figure}

These data have been used to search for a possible accumulation of events in the sky~\cite{IC9ps}. The resulting sky-average point-source sensitivity for a source with an $E^{-2}$ spectrum is  $E^2 d\phi/dE = 12 \times 10^{-8}~\rm GeVcm^{-2}s^{-1}$ which is already comparable with what was obtained with the 5 years of AMANDA II data presented in section~\ref{AMANDA}. The events were treated with a likelihood based analysis that makes use of the angular distribution of the background with a source hypothesis compared to a background only hypothesis obtained by scrambling the data in right ascension. The first significance map obtained with IceCube for the Northern hemisphere sky is shown on fig.~\ref{fig:IC9sky}. This map doesn't show any significant deviation from uniformity. The most significant excess, with a 3.3 $\sigma$ significance is at r.a. = 276.6$^{\circ}$, dec.= 20.4$^{\circ}$. This is comparable with a random fluctuation of a uniform background as 60\% of the datasets scrambled in right ascension show an excess of 3.3 $\sigma$ or higher somewhere in the sky. A search for neutrinos coming from 26 galactic and extragalactic preselected objects has also been performed on this dataset. In addition, the most significant excess over the expected background on these sources was found at the Crab nebula with 1.77$\sigma$, which again is consistent with random fluctuations. \\
Like in AMANDA, IceCube data can be used to probe the diffuse flux of neutrino from an unresolved population of astrophysical sources~\cite{diffuseIC9}. The sensitivity of IC-9 is $1.4 \times 10^{-7}~\rm GeV cm^{-2}s^{-1}sr^{-1}$ which is only a factor of 2 above the AMANDA-II sensitivity despite the much shorter integrated exposure time. Large improvements can be expected from both longer operation of IceCube with even more strings and refinement of analysis techniques.

\section{Conclusion : The future of IceCube}
\subsection{The next years of IceCube} 

The accumulated exposure of the IceCube 9 strings configuration does not allow us yet to reach the integrated exposure level required to probe astrophysical neutrino signals. Nevertheless, various analyses are developing and are very promising~\cite{IC9Atm}. These results confirm the stability of data taking, the good quality of the data recorded and experiment simulation. During the coming years, IceCube will continue to grow and will in 2009 reach an integrated exposure of 1 km$^3\cdot$yr. This will be an important milestone as it represents roughly what is needed to reach the level of detection for an  astrophysical neutrino flux~\cite{WB}. When completed, the acceptance of the detector will naturally be larger, but it will also have an improved performance for reconstruction due to its larger size. In the case of the search for point sources for instance, the longer lever arm for the reconstruction of the muons tracks will lead to a better angular resolution of the detector which will become better than a degree.

\subsection{One step further: extension of the IceCube detector at low energies with DeepCore} \label{DeepCore}
The capabilities of IceCube will be extended at both  lower and higher energies in the near future. Starting next austral summer, a compact core of 6 strings using IceCube's DOM technology, called the DeepCore detector, will start to be deployed near the center of the main in-ice detector. The interstring spacing will be of the order of 72~m, allowing for the exploration of energies as low as 10-20 GeV. The surrounding IceCube strings will be used as an active veto to reduce the atmospheric muon background. The energy range that is explored is very important for dark matter seaches that were initiated with AMANDA. Moreover, the ability to select contained events opens the search for downgoing astrophysical neutrino signals at low energies. This will allow one to look above the current horizon of IceCube, even opening the possibility to look at the galactic center or sources like RX J1713.7-3946~\cite{RXJ1713}.  

\subsection{The second step: extensions at higher energies}

At EeV energies, on the other end of the energy range, an extension of IceCube is also studied. The radio or the acoustic signal generated by neutrino interacting in the ice can be detected with a high energy extension of IceCube. With a much increased detection volume, we will aim at detecting the GZK neutrino flux. With attenuation lengths of the order of the kilometer for acoustic (kHz frequency range) and for radio signals (MHz frequency range), a sparse instrumentation will suffice for this extention. Two projects are currently explored for this extension: AURA (Askarian Underice Radio Array) for the radio signal~\cite{aura} and SPATS (South Pole Acoustic Test Setup) for the acoustic signal~\cite{spats}. They are currently studying the polar ice and developing the hardware necessary for the building of a hybrid detector enclosing IceCube in another array of strings with a much larger spacing that will allow to study these very scarce and energetic events.

\section*{Acknowledgments}
We acknowledge the support from the following agencies: National Science Foundation-Office of Polar
Program, National Science Foundation-Physics Division, University of Wisconsin Alumni Research Foundation,
Department of Energy, and National Energy Research Scientific Computing Center (supported by
the Office of Energy Research of the Department of Energy), the NSF-supported TeraGrid system at the San
Diego Supercomputer Center (SDSC), and the National Center for Supercomputing Applications (NCSA);
Swedish Research Council, Swedish Polar Research Secretariat, and Knut and Alice Wallenberg Foundation,
Sweden; German Ministry for Education and Research, Deutsche Forschungsgemeinschaft (DFG),
Germany; Fund for Scientific Research (FNRS-FWO), Flanders Institute to encourage scientific and technological
research in industry (IWT), Belgian Federal Office for Scientific, Technical and Cultural affairs
(OSTC); the Netherlands Organisation for Scientific Research (NWO);
M. Ribordy acknowledges the support of the SNF (Switzerland); A. Kappes and J. D. Zornoza acknowledge support by the EU
Marie Curie OIF Program.

\end{document}